\documentclass[twocolumn,showpacs,showkeys,preprintnumbers,amsmath,amssymb,aps]{revtex4}

\usepackage{graphicx}

\begin{document}

\preprint{MMM, Wurmehl, GD-01}

\title{Half-metallic ferromagnetism with high magnetic moment and high Curie temperature in Co$_2$FeSi}

\author{Sabine Wurmehl, Gerhard H. Fecher, Vadim Ksenofontov,
        Frederick Casper, Ullrich Stumm, and Claudia Felser}
\affiliation{Institut f\"ur Anorganische Chemie und Analytische Chemie,
             Johannes Gutenberg - Universit\"at,
             D-55099 Mainz, Germany}

\author{Hong-Ji Lin}
\affiliation{National Synchrotron Radiation Research Center, Hsinchu 30077, Taiwan}

\author{Yeukuang Hwu}
\affiliation{Institute of Physics, Academia Sinica, Taipei 11529, Taiwan}

\date{\today}

\begin{abstract}
Co$_2$FeSi crystallizes in the ordered L2$_1$ structure as proved by
X-ray diffraction and M\"o\ss bauer spectroscopy. The magnetic
moment of Co$_2$FeSi was measured to be about $6\mu_B$ at 5K.
Magnetic circular dichroism spectra excited by soft X-rays (XMCD)
were taken to determine the element specific magnetic moments of Co
and Fe. The Curie temperature was measured with different methods to
be ($1100\pm20$)K. Co$_2$FeSi was found to be the Heusler compound
as well as the half-metallic ferromagnet with the highest magnetic
moment and Curie temperature.
\end{abstract}



\maketitle

\section{Introduction}

Heusler compounds are ternary intermetallics with the composition
X$_2$YZ \cite{HEU03}. They order in the L2$_1$ structure, space
group $F\:m\overline{3}m$. The Co$_2$ based Heusler compounds are of
particular interest due to their high Curie temperatures and the
high magnetic moments per unit cell (for a review see Ref.
\cite{LB19C, LB32C}).

Half-metallic ferromagnets (HFM) are predicted to be 100\% spin
polarized \cite{GME83}. In those materials, the majority electrons
are metallic whereas the minority electrons are semi-conducting or
insulating. A high degree of spin polarization is interesting for
various applications in magneto-electronics (for examples see
\cite{CVB02}). Several materials in the class of Heusler compounds
were predicted to be half-metallic ferromagnets
\cite{GME83,IFK95,GDP02}.

This work reports about the Heusler compound Co$_2$FeSi that is
found to be, at present, the Heusler compound exhibiting the highest
magnetic moment as well as the highest Curie temperature.

\section{Results and Discussion}

Co$_2$FeSi samples were prepared by arc-melting of stochimetric
quantities of the pure metals in an argon atmosphere, followed by
annealing in sealed quartz tubes at 1300K for 21 days. The crystal
structure of the polycrystalline ingots was investigated by X-Ray
diffraction and proved to exhibit the correct L2$_1$ structure. The
lattice constant was found to be 5.64\AA by Rietveld refinement. The
Rietveld refinement allowed for a B2 type disorder of only $<10$\%
(mixing between Fe and Si). $^{57}$Fe M\"o\ss bauer spectroscopy was
performed to gain additional structural information. The derived
spectrum (not shown here) exhibited a single sextet consisting of
sharp lines with a width of 0.15mm/s being in the order of the
$\alpha$-Fe line width (0.136mm/s). This observation confirms the
occupation of a single site for Fe and thus a well ordered system.
DO$_3$ like disorder (mixing of Co and Fe at X and Y sites) can be
excluded from a comparison of the measured ($26.3\times10^6$A/m) and
calculated values ($21\times10^6$A/m for Fe in Y and
$11\times10^6$A/m for Fe in X positions) of the hyperfine field.

\begin{figure}[ht]
\centering
\includegraphics[width=8cm]{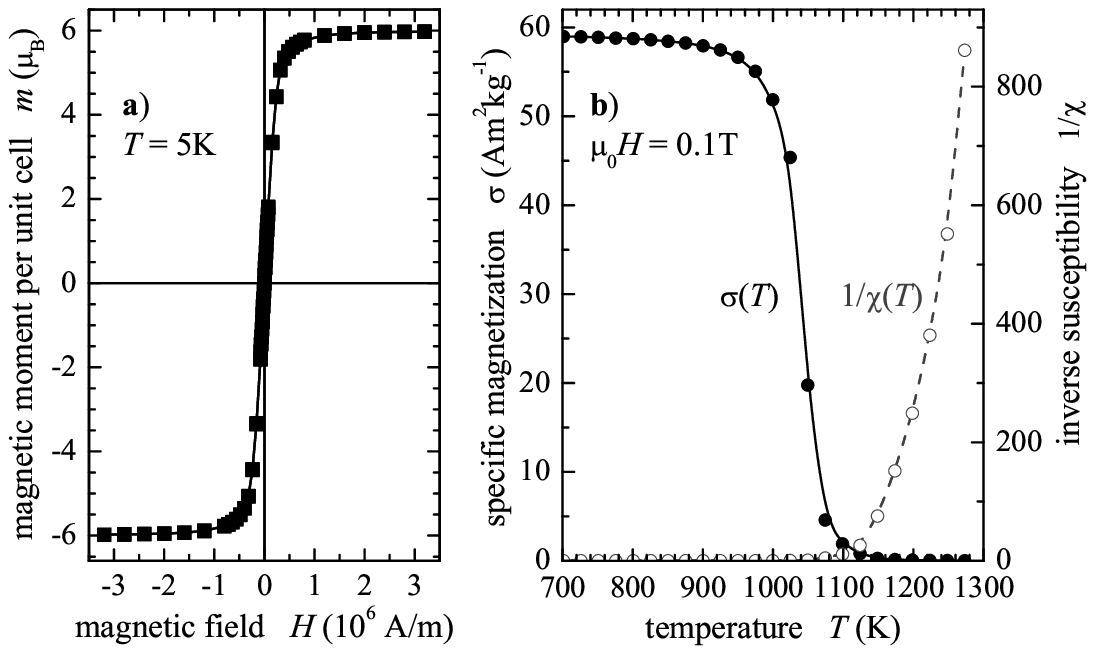}
\caption{Magnetization of Co$_2$FeSi.\\
         (a) shows the low temperature and (b) the high temperature magnetic properties
         as measured by SQUID and VSM, respectively.}
\label{fig1}
\end{figure}

Low temperature magnetometry was performed using a super-conducting
quantum interference device (SQUID). The results are displayed in
Fig.\ref{fig1}a.

The field dependence of the magnetization is typical for a
soft-magnetic material. The measured magnetic moment in saturation
is $(5.97 \pm 0.05)\mu_B$ at 5K resulting in $6\mu_B$ at 0K by
extrapolation. This value is in agreement with the spin moment
expected from the Slater - Pauling rule \cite{FKW05}. The measured
magnetic moment is an integer within the experimental uncertainty,
as expected for a half-metallic ferromagnet.

\begin{figure}[ht]
\centering
\includegraphics[width=8cm]{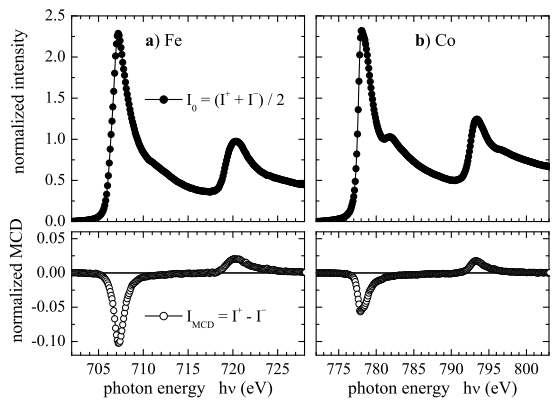}
\caption{Site resolved magnetic properties of Co$_2$FeSi.\\
         Shown are the XAS (I$_0$) and XMCD (I$_{MCD}$) spectra
         taken at the L$_{2,3}$ absorption edges of
         Fe (a) and Co (b) after subtracting a constant background.}
\label{fig2}
\end{figure}

X-ray Magnetic Circular Dichroism (XMCD) in photo absorption (XAS)
was measured at the {\it First Dragon} beamline of NSRRC (Hsinchu,
Taiwan). The XAS and XMCD spectra were taken at the L$_{2,3}$
absorption edges of Fe and Co. The results are shown in
Fig.\ref{fig2}.

The two white lines corresponding to the L edges are clearly seen
for Fe (a) and Co (b). An additional spectral feature is visible
3.5eV below the L$_3$ absorption edge of Co and a weaker one at the
Fe L$_3$ edge. This feature is related to the L2$_1$ structure and
demonstrates the high structural order of the sample (it vanishes
for B2 like disorder) \cite{con05}.

The magnetic moments per atom derived from a sum rule analysis
\cite{TCS92,CTA93} are $(2.6\pm0.1)\mu_B$ for Fe and
$(1.2\pm0.1)\mu_B$ for Co at $T=300K$ and $\mu_0H=0.4T$. The error
arises mainly from the unknown number of holes in the $3d$ shell and
the disregard of the magnetic dipole term in the sum rule analysis.
The orbital to spin magnetic moment ratios are about 0.05 for Fe and
0.1 for Co. The measured ratio between the Fe and Co spin moments of
2.1$\overline6$ is fully reproduced by the calculated ratio of 2.18.

A seeming linear dependence of the Curie temperature as a function
of the magnetic moment is observed in Co$_2$ based Heusler compounds
as shown in \cite{FKW05}. One expects that $T_C$ is highest for
those HMF exhibiting a large magnetic moment. $T_C$ should be above
1000K in Co$_2$FeSi with a magnetic moment of $6\mu_B$. Following
this suggestion, the ferromagnetic Curie temperature of Co$_2$FeSi
was measured with a vibrating sample magnetometer (VSM) equipped
with a high temperature stage. The result obtained in a constant
induction field of $\mu_0H=0.1T$ is shown in Fig.\ref{fig1}b. A
value of $(1100\pm20)K$ is obtained from the measurement.

The paramagnetic Curie-Weiss temperature $\Theta$ was estimated from
a plot of the inverse susceptibility (1/$\chi$) as a function of
temperature (see Fig.\ref{fig1}b). The Curie-Weiss temperature is
found by interpolating 1/$\chi(T)$ to be $(1150\pm50)$K. A true
linear behavior for 1/$\chi$ as a function of temperature is not
observed here because the experiment was performed in a temperature
range close to the Curie temperature. A linear dependence can be
expected from molecular field theory only for temperatures far above
$T_C$.

The here observed properties of Co$_2$FeSi are in agreement to those
reported previously by Niculescu {\it et al} \cite{NBR77,NBH79} for
a higher degree of disorder (10\% B2 plus 16\% DO$_3$). (Please
note: \footnote{Niculescu {\it et al} reported in \cite{NBR77}:
$a=5.66$\AA, $T_C>980$K and $m=5.9\mu_B$ at 10K; the lattice
parameter (5.64\AA) found here agrees with the one reported in
\cite{NBR77} for Fe$_2$CoSi; Values for $a$ and $m$ reviewed later
\cite{LB32C,Bus88} are considerably different, for unknown
reasons.}.)

\begin{figure}[ht]
\centering
\includegraphics[width=6.5cm]{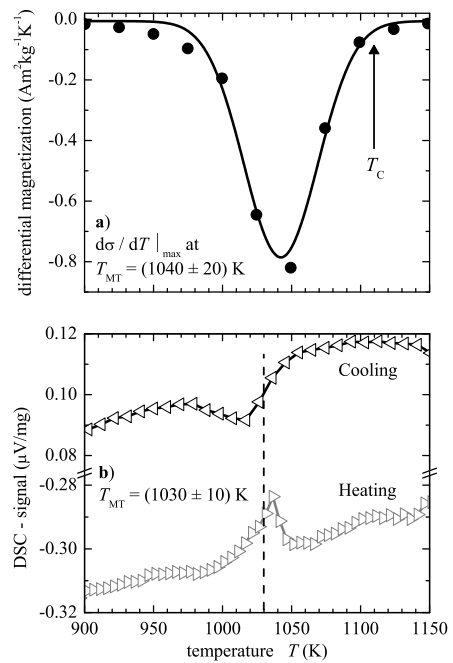}
\caption{Phase transition in Co$_2$FeSi. \\
         The differential magnetization (a) close to $T_C$ is compared to
         differential scanning calorimetry (b).}
   \label{fig3}
\end{figure}

SQUID magnetometry does oftenly not allow to determine high
temperature magnetic phase transitions. The low temperature
requirement of the instrumental set-up may not be met, in particular
if $T_C$ is very high. Differential scanning calorimetry (DSC) is
well established to investigate various kinds of phase transitions
in solid materials (for example see: \cite{KUF05}). Here it was used
to examine the magnetic phase transition in order to support the
Curie temperature received by the VSM experiments. Fig. \ref{fig3}
compares the DSC signal (b) with the derivative of the specific
magnetization (a) with respect to the temperature (compare
Fig.\ref{fig1}b). The minimum at $(1040\pm20)$K in Fig.\ref{fig3}a
corresponds to the maximum change of the magnetization with
temperature. With DSC, a pronounced shift of the signal is observed
during cooling or heating at about 1017K and 1037K, respectively.
This shift is due to the hysteresis of the DSC method and depends on
the temperature gradient $\dot{T}(t)$. The signals are attributed to
changes of the magnetic properties, as no structural transitions
were observed in this temperature range. Therefore, one expects a
mean value of $(1030\pm5)$K for the magnetic phase transition of
Co$_2$FeSi. The melting point was also observed by DSC and found to
be $T_m=(1517\pm5)$K (not shown in Fig.\ref{fig3}).

The observation of the magnetic transition by DSC is verified by the
comparison of the structures in DSC with the differential
magnetization. The latter is clearly observed at the point of
maximum change of the magnetization with temperature. It is seen
that the value obtained by DSC is slightly lower than $T_C$
determined from the VSM measurement. For Co$_2$FeSi one finds it to
be only 4.5\% below the Curie temperature and thus the DSC value may
be a simple estimate for $T_C$ if other methods are not available.

\section{Summary}

The structural and magnetic properties of the Heusler compound
Co$_2$FeSi were reported. Co$_2$FeSi has a lattice parameter of
5.64\AA and crystallizes in the L2$_1$ structure with very low
disorder. Its melting point appears at $(1517\pm5)$K. The material
is soft magnetic and its specific saturation magnetization of
0.166Am$^2$kg$^{-1}$ at 0K exceeds the one of pure Co by about 9\%.
The Curie temperature of 1100K is 5\% higher than that of pure Fe.
The magnetic moment of 6$\mu_B$ per unit cell points clearly on a
half-metallic ferromagnet.

As a practical application, it was shown how to estimate the lower
temperature limit of the magnetic phase transition for materials
with very high Curie temperatures by means of differential scanning
calorimetry.

\begin{acknowledgments}
This work is funded by the DFG (FG 559) and DAAD (D/03/31 4973).
\end{acknowledgments}

\end{document}